\documentclass{iau}
\usepackage{graphicx}

\usepackage{natbib}
\usepackage{abbrev}

\newcommand{\fdg}{\mbox{\ensuremath{.\!\!^\circ}}}
\newcommand{\arcdeg}{\ensuremath{^{\circ}}}

\title[Galactic Center Star Formation] 
{Star Formation in the Galactic Center}

\author[Jens Kauffmann]  
{Jens Kauffmann}

\affiliation{Max--Planck--Institut f\"ur Radioastronomie, Auf dem
  H\"ugel 69, 53125 Bonn, Germany\\
  email: \texttt{jens.kauffmann@gmail.com}}

\pubyear{2016}
\volume{322}  
\jname{The Multi--Messenger Astrophysics of the Galactic Centre}
\editors{R. M. Crocker, S. N. Longmore \& G. V. Bicknell, eds.}
\begin{document}

\maketitle

\begin{abstract}
Research on Galactic Center star formation is making great advances, in particular due to new data from interferometers spatially resolving molecular clouds in this environment. These new results are discussed in the context of established knowledge about the Galactic Center. Particular attention is paid to suppressed star formation in the Galactic Center and how it might result from shallow density gradients in molecular clouds.
\end{abstract}

\firstsection 

\section{Introduction}
The Galactic Center environment provides unique conditions for star formation within the Milky Way. About 3--10\% of the total molecular gas and star formation (SF) of the Milky Way\footnote{Observations of line and dust emission indicate that the $|\ell|\le{}3^{\circ}$ region contains of order $(3~{\rm{}to}~8)\times{}10^7\,M_{\odot}$ of molecular gas \citep{dahmen1998:cmz-gas, tsuboi1999:cs-nobeyama, longmore2012:sfr-cmz}. It forms stars at a rate of order $0.1\,M_{\odot}\,\rm{}yr^{-1}$ \citep{yusef-zadeh2009:sf-cmz, immer2012:recent-sfr, longmore2012:sfr-cmz, koepferl2015:masquerading-cmz}. The Milky Way contains $(1.0\pm{}0.3)\times{}10^9\,M_{\odot}$ of molecular gas \citep{heyer2015:review}, while the star formation rate is $1~{\rm{}to}~4\,M_{\odot}\,\rm{}yr^{-1}$ \citep{diehl2006:26al, misiriotis2006:ism-3d, lee2012:wmap}.} reside at $|\ell|\le{}3^{\circ}$ (i.e., within galactocentric radii $\le{}430~\rm{}pc$). Unlike the rest of the the Milky Way, the region within $\sim{}200~\rm{}pc$ galactocentric radius (i.e., $|\ell|\lesssim{}1\fdg{}5$) is dominated by gas in molecular instead of atomic form. This domain of the Galaxy is therefore also known as the Central Molecular Zone (CMZ; \citealt{morris1996:cmz-review}). The molecular clouds in the CMZ are unusually warm, dense, and ``turbulent'' (see below).

This text largely ignores the immediate environment of Sgr~A$^{\ast}$ and the question why the CMZ is so unusual within the Milky Way. Kruijssen et al.\ (this volume) discuss this latter aspect of CMZ science.  Instead this text sets out to summarize our knowledge about ongoing star formation and the state and distribution of dense molecular gas in the CMZ. Many of the most recent results concern the internal structure of CMZ clouds as resolved by interferometers. Many of the global aspects described in the reviews by \citet{gusten1989:review} and \citet{morris1996:cmz-review} thus still remain valid today. A Galactic Center distance of $(8.34\pm{}0.16)~\rm{}kpc$ is adopted \citep{reid2014:bessel}.

\section{Gas and Young Stars in the Galactic
  Center\label{sec:gas-stars}}
\subsection{Observations of Young Stars and Star Formation\label{sec:stars}}
The CMZ is a spectacular star--forming environment. It contains massive and compact young stellar groups like the Arches and Quintuplet clusters \citep{nagata1990:quintuplet, nagata1995:arches, cotera1994:arches} that together contain $\gtrsim{}200$ O--type stars \citep{figer2004:cmz-clusters} and have few --- but still some --- counterparts in the disk of the Milky Way \citep{portegieszwart2010:young-massive-clusters}. The CMZ also harbors the Sgr~B2 molecular cloud that alone hosts 59 compact H\textsc{ii}~regions \citep{gaume1995:sgr_b2-i}.

Still, relatively few and poor overall constraints on the young stars inhabiting the CMZ exist. For example, \emph{Spitzer} can at best detect embedded young stars with luminosities $\gtrsim{}10^3\,L_{\odot}$ in the CMZ (Eq.~[2] of \citealt{dunham2008:vellos}, degraded by a factor 100 given lower completeness in the CMZ taken from \citealt{evans2007:c2d-delivery} and \citealt{gutermuth2015:mipsgal}). But some CMZ clouds have high optical depths even at wavelengths $\sim{}100~\rm{}\mu{}m$ (e.g., G0.253+0.016; \citealt{lis1998:m025}). In these cases embedded star formation does at best manifest in subtle trends affecting the spectral energy distribution of the entire cloud \citep{lis2001:ir-spectra}. Similarly, OB--type stars outside clouds, which provide information on star formation during the past few $10^6~\rm{}yr$, are hard to detect behind foreground extinction at the level of $A_{K_{\rm{}s}}\approx{}2~\rm{}mag$ \citep{schodel2010:near-ir, longmore2011:m025}, corresponding to $A_V\approx{}18~\rm{}mag$ (for $A_K/A_V=0.113$; \citealt{rieke1985:extinction-law}).

%
%

The available data provide interesting insights, though. \citet{yusef-zadeh2009:sf-cmz} identify about 550 candidate young stellar objects from \emph{Spitzer} images at 8 and $24~\rm{}\mu{}m$ wavelength. \citet{immer2012:recent-sfr} use a similar but slightly refined technique to identify 1,141 candidate young stars in \emph{ISO} and \emph{MSX} data. These infrared observations yield star formation rates $\approx{}0.1\,M_{\odot}\,\rm{}yr^{-1}$ at at $|\ell|\le{}1\fdg{}5$. The nature of these objects is, however, not entirely clear. \citet{yusef-zadeh2009:sf-cmz} in particular point out that their objects have spectral energy distributions consistent with being deeply embedded in molecular clouds. But the \citeauthor{yusef-zadeh2009:sf-cmz} sources are at the same time found to typically reside away from the clouds that could envelope young stars. This produces an unclear picture. \citet{koepferl2015:masquerading-cmz} suggest that main sequence stars ``illuminating'' diffuse CMZ gas could produce infrared signals resembling those of embedded young stars. This complication has the potential to massively reduce the value of infrared CMZ data.

Also, Paschen--$\alpha$ imaging of the CMZ reveals an extended population of high--mass stars (\citealt{wang2010:cmz}; \citealt{dong2011:cmz-hst, dong2012:multiwavelength-cmz}). Interestingly, these stars might well have formed as part of the Arches and Quintuplet clusters but then migrated to their current location \citep{habibi2014:isolated-stars}.

The best constraints on the total star formation activity of the CMZ might therefore come from indirect methods. For example, the number of ionizing photons produced in the CMZ can be estimated from radio data (e.g., \citealt{mezger1979:hii-cmz} for some early research). This in turn constrains the number of high--mass stars emitting such photons, which itself can be related to the star formation rate via further assumptions. Early work (e.g., \citealt{gusten1989:review}: 0.3 to $0.6\,M_{\odot}\,\rm{}yr^{-1}$ at $|\ell|\le{}3\fdg{}5$) is broadly consistent with current estimates (e.g., \citealt{longmore2012:sfr-cmz}: $\le{}0.06\,M_{\odot}\,\rm{}yr^{-1}$ at $|\ell|\le{}1\arcdeg$). Section~\ref{sec:suppressed-sf} describes how water and methanol masers can be used for similar but more uncertain estimates.

\subsection{State and Distribution of Dense Gas\label{sec:gas}}
The distribution of molecular gas in the Galactic Center at $|\ell{}|\le{}5\arcdeg$ is highly asymmetric. About 75\% of the gas seen in $\rm{}^{13}CO$ and CS resides at $\ell{}>0\arcdeg$ and radial velocities $>0~\rm{}km\,s^{-1}$ \citep{bally1988:maps-cmz}. Interestingly, this distribution is opposite to potential asymmetries in star formation that are possibly seen in the aforementioned \emph{Spitzer} data \citep{yusef-zadeh2009:sf-cmz}. It is likely that the gas seen on these large spatial scales follows so--called $x_1$~orbits that are closed and elongated along the Milky Way's bar \citep{contopoulos1977:orbits, binney1991:cmz-orbits}. These orbits are also dynamically stable down to some minimum size.

Provided orbits are chiefly controlled by a bar--like potential, stable trajectories on spatial scales below those of $x_1$~orbits are part of the family of $x_2$~orbits. These are closed and elongated \emph{perpendicular} to the bar. Interestingly, \citet{molinari2011:cmz-ring} argue that the gas within $|\ell|\lesssim{}1\arcdeg$ from the Galactic Center forms a system of unusually dense and massive molecular clouds with kinematics that are broadly consistent with those of gas on $x_2$~orbits. These clouds include regions like Sgr~B2, G0.253+0.016, and all the other objects that form the main topic of this text.

However, \citet{henshaw2016:cmz-kinematics} recently demonstrated that the kinematics of clouds at $|\ell|\lesssim{}1\arcdeg$ are much better explained by the \emph{open} orbits \citet{kruijssen2014:orbit} propose for an azimuthally symmetric CMZ potential taken from \citet{launhardt2002:cmz-potential}. These trajectories are \emph{not stable}: the orbits can form intersection points where energy can be consumed in strong shocks. The orbits are not necessarily occupied by continuous streams of gas, though, and they oscillate (and thus potentially avoid another) perpendicular to the Galactic Plane. Numerical orbit simulations like those by \citet{lucas2015:thesis} indicate that gas can reside on these orbits for several 10~Myr. The orbital period of major CMZ clouds like Sgr~B2 is about 2~Myr in radius and 4~Myr in azimuth \citep{kruijssen2014:orbit}.

\citet{binney1991:cmz-orbits} suggest that $x_1$~orbits shrinking below the size of stable trajectories eventually shock and dump their material onto the inner CMZ (e.g., $x_2$~orbits). \citet{lucas2015:thesis} shows that the injection of a single compact cloud into the central $\sim{}100~\rm{}pc$ produces ``streams'' similar to those considered by \citet{kruijssen2014:orbit}.\medskip

\noindent{}CMZ molecular clouds have unusually high mean $\rm{}H_2$ densities $\sim{}10^4~\rm{}cm^{-3}$ and column densities $\sim{}10^{23}~\rm{}cm^{-2}$ (e.g., \citealt{lis1994:dust-ridge}). The diffuse ionized gas is pervaded by a strong magnetic field $\sim{}10^3~\rm{}\mu{}G$ \citep{yusef-zadeh1984:non-thermal, uchida1985:cmz-lobes,chuss2003:cmz-polarization, novak2003:cmz-polarization} that also penetrates the CMZ clouds \citep{pillai2015:magnetic-fields}.

CMZ molecular clouds have line widths much in excess of Galactic Disk clouds (Sec.~\ref{sec:kinematics}).  Many CMZ clouds appear to be subject to violent processes like cloud--cloud collisions at high velocities. This is indicated by widespread emission from SiO \citep{martin-pintado1997:sio-cmz, huettemeister1998:shocks, riquelme2010:survey} and other molecules likely ejected from grain surfaces via shocks \citep{requena-torres2006:corganic-mols, requena-torres2008:coxygen-coms}, and methanol masers excited in collisions (\citealt{mills2015:cmz-masers}; also see \citealt{menten2009:g1.6}, though).

Bulk gas temperatures from line ratios are typically in the range 50 to $100~\rm{}K$ (\citealt{guesten1981:nh3-cmz, ao2013:cmz-temperatures, ott2014:cmz-atca, ginsburg2015:cmz-gas-temperatures}; also see \citealt{riquelme2010:isotopes, riquelme2012:temp-loop-interceptions}). \citet{huettemeister1993:nh3-cmz} do, however, point out that the $\rm{}NH_3$--derived temperature in a given cloud varies between $\approx{}25~\rm{}K$ and $\gtrsim{}200~\rm{}K$ (also see \citealt{mills2013:widespread-hot-nh3}), and that the cold material contains $\sim{}75\%$ of the mass traced by $\rm{}NH_3$. This topic should be revisited systematically. It has been suggested that temperatures from some line ratios are unphysically high due to formation pumping \citep{lis2014:h3o+}, but this requires densities below the few $10^4~\rm{}cm^{-3}$ characteristic of the CMZ (D.~Lis, priv.\ comm.). Gas temperatures $\gtrsim{}50~\rm{}K$ would be mysteriously decoupled from the much lower dust temperatures $\approx{}20~\rm{}K$ (e.g., \citealt{guesten1981:nh3-cmz}, \citealt{molinari2011:cmz-ring}, \citealt{longmore2011:m025}). This could be explained if gas was heated by agents not affecting the dust, such as cosmic rays (see \citealt{clark2013:gas-dust-g0253} for recent modeling work).

The heating of CMZ gas is an unsolved problem, though. \citet{ginsburg2015:cmz-gas-temperatures} point out that temperatures even vary \emph{within} given CMZ clouds. Heating via cosmic rays, however, should provide relatively homogeneous heating throughout the clouds. \citeauthor{ginsburg2015:cmz-gas-temperatures} therefore conclude the gas is chiefly heated by turbulence. \citet{immer2016:cmz-temperature} indeed find a strong correlation between gas temperature and line width supporting this picture.

\section{Conditions for Star Formation: Updates \& Personal Insights\label{sec:processes}}
\subsection{High Gas Temperatures imply large Bonnor--Ebert Masses}
Here I discuss issues that are insufficiently explored elsewhere or that deserve highlighting due to new results. The former includes the application of the analysis by \citet{ebert1955:be-spheres} and \citet{bonnor1956:be-spheres} to the CMZ. They show that cloud fragments must exceed a mass
\begin{equation}
m_{\rm{}BE} = 20 \, M_{\odot} \cdot (T_{\rm{}gas}/50~{\rm{}K})^{3/2} \cdot (n_{\rm{}H_2}/10^5~{\rm{}cm^{-3}})^{-1/2}
\end{equation}
before they become unstable to gravitational collapse. This threshold mass is remarkably large in the Galactic Center: gas temperatures $T_{\rm{}gas}$ in the CMZ exceed values found closer to Sun by a factor $\sim{}5$, so that $m_{\rm{}BE}$ is larger by a factor $\sim{}5^{3/2}\approx{}11$ in Galactic Center clouds for fixed density $n_{\rm{}H_2}$. This suggests that star--forming cloud fragments in the CMZ are unusually massive (which might promote the formation of high--mass stars) or dense compared to regions elsewhere in the Milky Way.

\subsection{Radial Tidal Forces are not destructive}
It is often stated that only CMZ molecular clouds with densities $\gtrsim{}10^4~{\rm{}cm^{-3}}\cdot{}(r_{\rm{}GC}/75~{\rm{}pc})^{-1.8}$ can survive Galactic Center tides \citep{gusten1989:review}. Current calculations, however, show that clouds at galactocentric radii of $20~{\rm{}to}~100~\rm{}pc$ are generally subject to \emph{compressive} tidal forces in the radial direction (e.g., Fig.~6.2 of \citealt{lucas2015:thesis}) because the gravitational force $F_{\rm{}g}\propto{}m/r_{\rm{}GC}^2$ \emph{increases} with increasing galactocentric radius for the observed CMZ mass profile, $m\propto{}r_{\rm{}GC}^{2.2}$ (\citealt{launhardt2002:cmz-potential}; see \citealt{kruijssen2014:orbit} for the power--law). This updates the classical discussion by \citet{gusten1980:h2co-cmz} who assume $m\propto{}r_{\rm{}GC}^{1.2}$.

\subsection{High Gas Densities from Confining Pressure}
X--Ray images do since a long time hint at the existence of hot and tenuous CMZ gas at high pressure \citep{yamauchi1990:cmz-plasma, spergel1992:pressure-bulge, muno2004:diffuse-xray}. \citet{ponti2015:cmz-xmm} suggest that candidate supernova remnants at pressure $\sim{}5\times{}10^6~\rm{}K\,cm^{-3}$ blow shells into the gas throughout the CMZ. We may assume that CMZ molecular clouds are in balance with a lower but still similar pressure. Given gas temperatures $T_{\rm{}gas}\approx{}50~\rm{}K$,
\begin{equation}
P/k_{\rm{}B} = 5\times{}10^5~{\rm{}K\,cm^{-3}} \cdot
(T_{\rm{}gas}/50~{\rm{}K}) \cdot
(n_{\rm{}H_2}/10^4~{\rm{}cm^{-3}})
\end{equation}
(where $k_{\rm{}B}$ is the Boltzmann constant) then implies high gas densities. This might be the chief reason why all CMZ clouds are indeed observed to have densities $n_{\rm{}H_2}\gtrsim{}10^4~\rm{}cm^{-3}$.

\section{Recent Results: Star Formation in Dense Galactic Center Clouds\label{sec:recent-insights}}

\subsection{Improved Assessments of Suppressed Star Formation in Dense Gas\label{sec:suppressed-sf}}
Given the high gas densities of CMZ clouds, one particularly surprising feature of the region is that \emph{star formation in the dense gas of the CMZ appears to be suppressed}. \citet{taylor1993:cmz-water-masers} concludes that, given massive and dense clouds, the CMZ should contain about an order of magnitude more $\rm{}H_2O$ masers than observed (following \citealt{guesten1983:h2o-masers} and \citealt{caswell1983:water-masers}; see \citealt{caswell1996:masers-methanol} for methanol masers). Observations with increased angular resolution eventually revealed individual clouds with little star formation (\citealt{lis1994:m0.25, lis2001:ir-spectra}; \citealt{lis1998:m025}; also see Sec.~\ref{sec:density-structure}).

Our ability to quantify the relation between star formation and dense gas has improved massively over the last few years. Several studies focusing on the Solar Neighborhood \citep{heiderman2010:sf-law, lada2010:sf-efficiency, evans2014:sfr-nearby-clouds} provide a framework against which Galactic Center clouds can be compared. Characterizations of the dense gas and star formation activity in the CMZ are not straightforward, as already highlighted in Sec.~\ref{sec:stars}. The degree--scale assessments of star formation in the CMZ by \citet{longmore2012:sfr-cmz} build on estimates for the number of ionizing photons from WMAP data. Studies of individual clouds (e.g., \citealt{kauffmann2013:g0.253, kauffmann2016:gcms_i, kauffmann2016:gcms_ii}) use data on H\textsc{ii}~regions and class--II methanol masers embedded in clouds. This yields an estimate of the number of embedded high--mass stars, which in turn hints at the star formation rate. In future assessments of the embedded population of $\rm{}H_2O$~masers provides an alternative way to gauge the star formation rate (e.g., \citealt{lu2015:20kms}; also see this volume). Characterizing the dense gas is even more difficult. Following \citet{lada2010:sf-efficiency}, one may broadly consider material at $A_V\gtrsim{}7~\rm{}mag$ to be ``dense''. Most current CMZ studies use this criterion to identify dense material on the basis of column density maps from, e.g., dust emission. \citet{longmore2012:sfr-cmz} also explore an approach in which all $\rm{}NH_3$--emitting gas is considered to be dense.

In summary, analysis shows that CMZ clouds are by about an order of magnitude less efficient in producing stars out of dense gas than clouds closer to Sun. This holds for averages over degree--sized parts of the region \citep{longmore2012:sfr-cmz} as well as for individual clouds \citep{kauffmann2013:g0.253}. We need to develop a detailed understanding of the conditions in the CMZ in order to understand this suppression of star formation.

\subsection{Density Structure of Molecular Clouds\label{sec:density-structure}}
Our understanding of the internal structure of CMZ molecular clouds has increased massively over the last few years. The cloud G0.253+0.016 might serve as an example. This region was first discovered in $\rm{}NH_3$ maps of the CMZ \citep{guesten1981:nh3-cmz}. Imaging of the CMZ in dust emission later revealed first details about the cloud structure \citep{lis1994:dust-ridge}. \citet{lis1994:m0.25} realized at this point that G0.253+0.016 is very extreme in its star formation properties: the cloud concentrates a mass resembling the one of the Orion~A molecular cloud in just $\sim{}3~\rm{}pc$ radius --- but there is no signifiant star formation in this object. A single faint $\rm{}H_2O$ maser, such as expected in regions of low--mass star formation, is the only signpost indicating that young stars exist in this cloud. Subsequent single--dish work by \citet{lis1998:m025} and \citet{lis2001:ir-spectra} further refined the properties of the cloud and its star formation activity. Infrared observations of G0.253+0.016 taken during this period inspired \citet{egan1998:irdcs} and \citet{carey1998:irdc-properties} to coin the term ``Infrared Dark Cloud'' (IRDC) for regions opaque at wavelengths $\gtrsim{}8~\rm{}\mu{}m$. Research on CMZ clouds then stopped for several years, given instrumental limitations. \citet{longmore2011:m025} revived this line of work with a fresh look at the object (now a.k.a.\ the ``Brick'') that is primarily motivated by new data from \emph{Herschel}.

None of the aforementioned studies did, however, resolve the internal structure of CMZ clouds. This is a problem: single--dish data probing spatial scales $\gtrsim{}1~\rm{}pc$ constrain how dense molecular cores capable of star formation aggregate out of the diffuse cloud medium. The observations do, however, not reveal the cores themselves on spatial scales $\lesssim{}0.1~\rm{}pc$ where individual stars form. This means that no constraints on the immediate initial conditions for CMZ star formation can be obtained.

Interferometer observations spatially resolving CMZ clouds constitute one of the major recent advances in research exploring Galactic Center star formation. A first study of G0.253+0.016 with the \emph{Submillimeter Array} (\emph{SMA}) by \citet{kauffmann2013:g0.253} reveals a puzzling trend: the cloud has a very high average density, that e.g.\ exceeds that of the Orion~A cloud by an order of magnitude, but \emph{the cloud is essentially devoid of significant dense cores} with radii $\lesssim{}0.1~\rm{}pc$. This trend essentially manifests in rather faint detections of $\rm{}N_2H^+$ in \emph{SMA} maps and an absence of significant dust continuum emission in the data. This result is not a consequence of a low sensitivity of the \emph{SMA} data: cloud cores resembling Orion~KL but located in the CMZ, for example, would be easily detected in such data. Even more detail in G0.253+0.016 is revealed by the \emph{ALMA} data of
\citet{rathborne2014:g0253-pdf, rathborne2015:g0253-alma}. Their dust emission maps confirm the absence of dense cores resembling Orion~KL, i.e., the relative absence of significant dense cores. Similarly, they also show that probability density functions (PDFs) of column density are devoid of excesses at high column density. This is typical for clouds with little star formation activity \citep{kainulainen2009:column-density-pdf} and quantifies that the cloud is not efficient in concentrating mass at high density.

A variety of interferometer--based studies of CMZ clouds have been published in the meantime. This includes further \emph{SMA} work on G0.253+0.016 \citep{johnston2014:g0.253}, studies of Sgr~C \citep{kendrew2013:sgr-c}, and research into the $20~\rm{}km\,s^{-1}$ cloud (\citealt{lu2015:20kms}; also see this volume). Further work on clouds in the so--called ``dust ridge'' is conducted by Walker et al.\ (this volume; also see \citealt{walker2015:dust-ridge}). At the same time \citet{immer2012:multi-wavelength-cmz} used the VLA to search for faint H\textsc{ii}~regions embedded in dust ridge clouds. They find little star formation in this region, but \citet{rodriguez2013:g0.253} use the same data to identify several new compact sources near and inside G0.253+0.016. \citet{mills2015:cmz-masers} also detect these objects, but they conclude that these are spatially extended features that are not consistent with being embedded H\textsc{ii}~regions.

The aforementioned work gives us a good idea of the spatially resolved properties in a few CMZ clouds. What is now needed is a comprehensive interferometric survey that covers most or all of the CMZ. Battersby et al.\ (this volume) present first results from the CMZoom project that uses the \emph{SMA} to develop such an overview. In the meantime the Galactic Center Molecular Cloud Survey (GCMS; \citealt{kauffmann2016:gcms_i, kauffmann2016:gcms_ii}) provides an \emph{SMA}--based overview of the resolved properties in all major CMZ clouds (i.e., Sgr~C, $20~\rm{}km\,s^{-1}$ cloud, $50~\rm{}km\,s^{-1}$ cloud, G0.253+0.016, and Sgr~B1--off). In addition, the GCMS already includes further \emph{ALMA} observations of G0.253+0.016 and selected fainter CMZ clouds (Kauffmann et al., in prep.). Also, guaranteed GCMS observations mean that we will possess \emph{ALMA} data for all major CMZ clouds by the end of cycle~4.\medskip

\noindent{}The data produce a coherent picture of CMZ cloud structure. The GCMS sample of \citet{kauffmann2016:gcms_ii}, e.g., shows that many CMZ clouds (i.e., $20~\rm{}km\,s^{-1}$ cloud, $50~\rm{}km\,s^{-1}$ cloud, G0.253+0.016, and Sgr~B2) have unusually flat density profiles resembling $\varrho{}\propto{}r^{-1.3}$. Only one cloud (i.e., Sgr~C --- plus Sgr~D, which probably resides outside the CMZ) has a steep density profile similar to a relation $\varrho{}\propto{}r^{-2}$ that would resemble the profiles expected in regions with ongoing star formation \citep{kauffmann2010:mass-size-ii}.

In other words, CMZ star formation is at least in part suppressed because CMZ clouds are inefficient in producing high--mass dense cores of size $\lesssim{}0.1~\rm{}pc$ that could efficiently produce stars. Section~\ref{sec:synthesis} discusses why CMZ clouds might have this structure.

\subsection{Kinematic Properties of Molecular Clouds\label{sec:kinematics}}
\citet{shetty2012:linewidth-size-cmz} show that the velocity dispersions $\sigma(v)$ in CMZ molecular clouds exceed those prevailing elsewhere in the Milky Way by a factor $\approx{}5$, when examined at a spatial scale $\sim{}5~\rm{}pc$. This trend has been known since the beginning of CMZ research (e.g., see \citealt{spergel1992:pressure-bulge} for an early compilation). This research did, however, not constrain the spatially resolved velocity field within clouds. This is unfortunate since gas kinematics control the cloud stability against self--gravity via the virial parameter
\begin{equation}
\alpha = 5 \, \sigma^2(v) \, R / (G \, M)
\end{equation}
(\citealt{bertoldi1992:pr_conf_cores}; $G$ is the Gravitational constant, while $M$ and $R$ are the cloud fragment's mass and radius). The new interferometer data add this critical information.

Interestingly, data on the comprehensive GCMS sample from \citet{kauffmann2016:gcms_i} confirm that CMZ clouds have unusually high velocity dispersions when analyzed on spatial scales $\gtrsim{}1~\rm{}pc$ --- \emph{but this excess appears to vanish on smaller spatial scales}. This trend had previously been found in the \emph{SMA} \citep{kauffmann2013:g0.253} and higher--quality \emph{ALMA} data \citep{rathborne2015:g0253-alma} for G0.253+0.016. In other words, the interferometer data now indicate that relatively narrow line widths $\le{}1~\rm{}km\,s^{-1}$ are found in dense cores throughout the CMZ. This implies an unusually steep CMZ linewidth--size relation.

Combination with the observed density structure indicates that CMZ clouds are gravitationally unbound on scales of a few parsec (i.e., $\alpha{}\gg{}2$: see \citealt{kauffmann2013:virial-parameter}), but that substructure on smaller spatial scales \emph{is} bound and potentially subject to gravitational collapse \citep{kauffmann2016:gcms_ii}.

\section{Synthesis: Star Formation Ability of Galactic Center Clouds\label{sec:synthesis}}
The discussion above can be summarized in three points. (\textit{i})~The star formation ability in the dense gas of CMZ clouds is suppressed by a factor $\sim{}10$, compared to regions closer to Sun. (\textit{ii})~High--mass dense cores capable of significant star formation are relatively absent in CMZ clouds. (\textit{iii})~The steep linewidth--size relation prevailing in the Galactic Center might mean that bound dense cores of size $\lesssim{}0.1~\rm{}pc$ are embedded in unbound clouds dominated by highly supersonic gas motions. The latter two factors are likely to influence star formation.

First, \citet{kruijssen2013:sf-suppression-cmz} argue (building on \citealt{krumholz2005:general-sf-theory} and \citealt{padoan2011:sf-supersonic-mhd}) that the supersonic gas motions prevailing in the CMZ massively increase the threshold density required for star formation. Analysis in \citet{kauffmann2016:gcms_ii} indeed gives threshold densities for star formation of $10^{7~{\rm{}to}~8}~\rm{}cm^{-3}$ in the CMZ, where \citet{kruijssen2013:sf-suppression-cmz} obtain densities $\sim{}10^4~\rm{}cm^{-3}$ for the Solar Neighborhood. This strongly suggests that supersonic turbulence is one of the factors suppressing CMZ star formation. Also see \citet{bertram2015:sf-efficiencies} on this point.

Note, however, that \emph{CMZ molecular clouds would actually exceed this threshold if they had a density structure typical for regions elsewhere in the Milky Way} (i.e., resembling $\varrho{}\propto{}r^{-2}$). This is, e.g., shown in Fig.~5 of \citet{kauffmann2016:gcms_ii}. For this reason it seems plausible to assume that \emph{the flat density structure of clouds is the chief factor suppressing CMZ star formation}: stars simply do not form because high--mass dense cores capable of significant star formation do not exist.

This, in turn, raises the question why such cores cannot form efficiently in the CMZ, i.e., why the density structure of CMZ clouds is unusually flat. We may speculate that shallow density gradients can emerge when the clouds are not tightly bound by self--gravity, so that gravity is not effective in building massive cores. The high line widths found in the CMZ would explain the low levels of gravitational binding.

The high gas velocity dispersions, finally, might be a consequence of the cloud--cloud collisions mentioned in Sec.~\ref{sec:gas}. These could be related to the potentially self--intersecting cloud orbits proposed by \citeauthor{kruijssen2014:orbit} (\citeyear{kruijssen2014:orbit}; also see \citealt{lucas2015:thesis}). Also see Kruijssen et al.\ (this volume) and \citet{krumholz2015:alpha-disk} on orbit dynamics.

Naturally, the strong magnetic field might also play a role in suppressing CMZ star formation. Observations of dust emission suggest the presence of fields with a strength $\approx{}5~\rm{}mG$ that cloud balance against self--gravity \citep{pillai2015:magnetic-fields}.

\section{Galactic Center Star Formation: A Template for
  Starbursts?\label{sec:template-starbursts}}
It is often said that CMZ might serve as a template for unresolved processes that are active in nearby and more distant starburst galaxies. For example, NGC~253 \citep{sakamoto2011:ngc253} and the Antennae Galaxies (NGC~4038/39; \citealt{wei2012:antennae}) contain molecular cloud complexes with mean $\rm{}H_2$ column densities $\sim{}10^{23}~\rm{}cm^{-2}$: these regions must be composed of clouds with column densities of the same order, i.e., clouds resembling those in the CMZ. See \citet{kruijssen2013:mw-vs-galaxies} for a related comparison between the Milky Way and other galaxies.

Still, extreme caution is required when using the CMZ as a template for the interpretation of other galaxies. The CMZ molecular clouds are, for example, apparently subject to high gas temperatures, cloud--cloud collisions, and orbital periods of just a few $10^6~\rm{}yr$. It is not clear that the evolution of these clouds will resemble those of regions in starbursts that are also warm but that reside on orbits with radii of a few $10^3~\rm{}pc$ and perturbation time scales $\gg{}10^6~\rm{}yr$. We need to disentangle all relevant processes before we can use CMZ clouds as templates to interpret the cosmos. That said, it is certainly instructive to see suppressed star formation in the CMZ. This clearly tells us that density is not the only factor controlling star formation.

\bibliographystyle{aa}
\bibliography{/Users/jens/texinputs/mendeley/library}

\end{document}